\documentclass{svjour3}                     
\usepackage{a4wide}
\usepackage{graphicx}
\usepackage{latexsym,amsbsy,amsmath}
\renewcommand{\vec}[1]{\boldsymbol{#1}}

\def\llcol#1#2{\tilde{\lambda}_{#1}.\tilde{\lambda}_{#2}}
\def\llss#1#2{\tilde{\lambda}_{#1}.\tilde{\lambda}_{#2}\,\boldsymbol{\sigma}_{#1}\boldsymbol{\sigma}_{#2}}
\begin{document}

\title{Fully heavy multiquarks%
 \thanks{Contribution to the 8th APFB Conference, Kanazawa (Japan), March 2021}}


 \author{Jean-Marc Richard}
%
%
 \institute{Jean-Marc Richard \at
              Universit\'e de Lyon, Institut des 2 Infinis de Lyon \\
               UCBL \& IN2P3
               4, rue Enrico Fermi\\
               F 69622 Villeurbanne\\
               \email{j-m.richard@ip2i.in2p3.fr}}
              

\date{Received: date / Accepted: date}

\maketitle

\begin{abstract}
The existence of stable multiquarks made  of heavy quarks and antiquarks is discussed in the framework of potential models. It is stressed that the few-body problem should be handled seriously and accurately. No bound state is found within the current models. But resonances are likely present in the spectrum above the lowest threshold. 
\end{abstract}

\section{Introduction}
\label{intro}
An interesting limit of hadron spectroscopy is that where all constituent quarks are heavy, and the light quarks contribute only by virtual pairs in the field surrounding the heavy quarks and bind them together. For some recent reviews about the exotics and the role of heavy quarks, with refs.~to earlier contributions, see, e.g., \cite{Brambilla:2019esw,Ali:2019roi,Richard:2016eis,Richard:2020uan}

There are, thus, reasonable expectations that the heavy quark limit is the realm of the Born-Oppenheimer regime: for a given position of each heavy quark, the gluon and light-quark energy is minimized, and this gives the effective potential in which the heavy quarks evolve. The question is whether or not the multiquark potential is pairwise and can be deduced from the quarkonium potential. 

The usual modeling consists of a pairwise interaction with the structure of a color-octet exchange, namely
\begin{equation}\label{eq:pot1}
 V=-\frac{3}{16}\sum_{i<j}\tilde\lambda_i.\tilde\lambda_j\,v(r_{ij})~,
\end{equation}
where $\tilde\lambda_i$ denotes the color generators. The normalization is such that $v(r)$ is the quarkonium potential. The central interaction \eqref{eq:pot1} can be supplemented by a spin-spin term
\begin{equation}\label{eq:pot2}
 V_\text{SS}=-\frac{3}{16}\sum_{i<j}\tilde\lambda_i.\tilde\lambda_j\,\vec\sigma_i.\vec\sigma_j\,\frac{k}{m_i\,m_j}\,\delta_\sigma(\vec r_{ij})~,
\end{equation}
where the $m_i$ are the quark msses and $\delta_\sigma$ is a smeared version of the contact interaction, with a short-range $\sigma$. This term, however, plays a smaller and smaller role when the quark masses increase. 

The model depicted in Eqs.~\eqref{eq:pot1} and  \eqref{eq:pot2} is of course rather crude, and,  for a $Q\bar Q$ meson or a $QQQ$ baryon, there is no point in computing the energy up to a fraction of keV. However, in multiquark spectroscopy, the first issue is whether or not such a simple dynamics leads to stable multiquarks (the next issue is the question of resonances, as discussed in the last section). Hence it is crucial to estimate the energy of ordinary hadrons and exotic hadrons consistently, with the same rigor. For instance, underestimating the 4-body energies by a substantial amount while the mesons are correctly calculated,  generates stable tetraquarks that do not actually exist in the model. 
\section{Tetraquarks}
\label{sec:tetra}
The earliest calculations based on \eqref{eq:pot1} concluded that configuration like $QQ\bar Q'\bar Q'$ cannot be stable unless the mass ratio $m(Q')/m(Q)$ is very large, actually much larger than $m_b/m_c$. Other configurations such as $bc\bar b\bar c$ are also unbound. See, e.g., \cite{Richard:2016eis,Richard:2018yrm,Richard:2020uan} and refs.\ there. 

These early estimates were based on genuine 4-body calculations. Several developments were published: first, improvements of the models, and refinements of the treatment of the 4-body dynamics; and also, unfortunately, attempts to simplify the modeling that were not always convincing. 

In particular,  the approximation of a point diquark artificially produces some extra attraction. Indeed,  replacing $v(r_{31})+v(r_{32})$ by $2\,v(r_{1a})$, where $a$ is the middle of particles $1$ and $2$,  is perfect for a Coulomb interaction if particles 1 and 2 are orbiting isotropically around each other: this is the Gauss theorem. For a linear interaction, this underestimates the interaction. For a quadratic potential, one obtains for a color $\bar 33$ configuration
\begin{equation}\label{eq:diquark}
 V{}=\frac12\left(r_{12}^2+r_{34}^2\right)
 +\frac34\sum_{i=1,2\atop j=3,4} r_{ij}^2
 {}=\frac34 (\vec x^2+\vec y^2)+3_,\vec z^2~,
\end{equation}
where
\begin{equation}
 \vec x=\vec r_2-\vec r_1~,\quad
 \vec y=\vec r_4-\vec r_3~,\quad
 \vec z=\frac{\vec r_3+\vec r_4-\vec r_1-\vec r_2}{2}~,
\end{equation}
is a standard set of Jacobi coordinates,
while a naive diquark approximation leads to  
$(\vec x^2+\vec y^2)/2+3\,\vec z^2$, which omits a positive contribution.  

One may wonder why in the pure chromoelectric limit \eqref{eq:pot1}, the equal-mass state $QQ\bar Q\bar Q$ is unstable, while the positronium analog Ps$_2(e^+e^+e^-e^-)$ is stable (when annihilation is disregarded). The authors of \cite{Karliner:2016zzc} ingenuously argue: ``The existence of dipositronium \dots implies that an analog di-quarkonium state exists''. This is not that simple, and, indeed, the empirical calculation of \cite{Karliner:2016zzc} finds only states above the threshold, in spite of the diquark approximation. 

\begin{figure}[ht!]
 \centering
 \includegraphics[width=.5\textwidth]{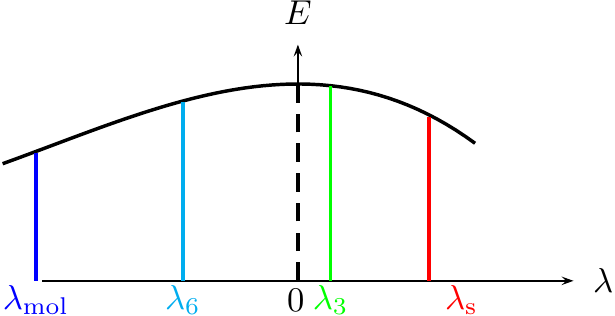}
 \caption{Four-body energy for two atoms ($\lambda_\text{s}$) vs.\ Ps$_2$  ($\lambda_\text{mol}$) or two mesons  ($\lambda_s$) vs.\ tetraquarks with either $\bar 33$ ($\lambda_3$) or $6\bar6$ ($\lambda_6$) color.}
 \label{fig:E-lambda}
\end{figure}

To analyze the difference between the dipositronium and the heavy-quark cases, consider the following  4-body Hamiltonians: the Ps$_2$ molecule with $v(r)=-1/r$; the $\mathrm{Ps}+\mathrm{Ps}$ threshold taken as a whole 4-body system; the $QQ\bar Q\bar Q$ tetraquark with color $\bar3 3$ and a purely central potential $v(r)$ of quarkonium type, as per \eqref{eq:pot1}; $QQ\bar Q\bar Q$ with color $6\bar 6$; the $Q\bar Q+Q\bar Q$ threshold seen as a 4-body system. In all cases, as simple rescaling leads to constituent masses $m=1/2$. With a suitable numbering, the above Hamiltonians can be recasted as 
\begin{equation}\label{eq:H-lambda}
H(\lambda)=\sum_{i=1}^4 \vec p_i^2 +\left(\frac13+2\,\lambda\right) (v_{12}+v_{34})+\left(\frac13-\lambda\right)(v_{13}+\cdots+v_{24})~,
\end{equation}
where $\lambda$ is a parameter measuring the asymmetry in the distribution of the potential enerrgy among the pairs, with values $\lambda=0$ for a fictitious set of four identical bosons, $\lambda_\text{s}=1/3$ for $\mathrm{Ps}+\mathrm{Ps}$ or $Q\bar Q+Q\bar Q$, $\lambda_\text{mol}=-2/3$ for Ps$_2$,  $\lambda_3=1/12$ for  $Q\bar Q+Q\bar Q$ in color $3\bar 3$, and $\lambda_6=-7/24$ for $QQ\bar Q\bar Q$ in color $6\bar 6$.   The ground state energy $E(\lambda)$ of $H(\lambda)$ is maximal for the symmetric case $\lambda=0$ (a mere consequence of the variational principle), and is decreasingly concave on both sides of $\lambda=0$, so that $0<\lambda<\lambda'$ implies rigorously $E(\lambda)>E(\lambda')$, and similarly for negative values.   Moreover, $E(\lambda)$  is likely to be nearly symmetric, so that one observes (but this is not fully rigorous) that  $0<|\lambda|<|\lambda'|$ implies $E(\lambda)>E(\lambda')$. In Fig.~\ref{fig:E-lambda} the energy $E(\lambda)$ is schematically drawn together with some relevant values of $\lambda$.

Clearly Ps$_2$ is favored, because the coefficients in \eqref{eq:H-lambda}, $\{-1,-1,+1,+1,+1,\discretionary{}{}{}+1\}$ have a wider spread around their mean value $1/3$, as compared to the threshold distribution $\{1,1,0,0,0,0\}$, i.e., $|\lambda_\text{mol}|>|\lambda_\text{s}$  Of course, the stability of Ps$_2$ is not explained exactly that way in the textbooks on Quantum Chemistry, but there is no contradiction: the wider the spread of the strength factors, the easier the possibility of polarizing the positronium atoms and inducing an attraction between them. 

On the other hand, none of the frozen color configurations $\bar 3 3$ and $6\bar 6$ is asymmetric enough to compete with the threshold, as $\lambda_3<\lambda_\text{s}$ and $|\lambda_6|<\lambda_\text{s}$. It is nevertheless understood while the latter is closer to binding. In numerical calculations, it is observed that the mixing of $6\bar 6$ with $\bar 3 3$ is not able to generate binding. 
\section{Heavy pentaquarks}
\label{se:pentaquarks}
The announcement by the LHCb collaboration of states interpreted as $\bar c cqqq$ (see, e.g., \cite{Brambilla:2019esw,Ali:2019roi}) stimulated studies revisiting pentaquarks with a single heavy quark or antiquark or extrapolating towards heavier states, up to $\bar QQQQQ$, where $Q$ denotes either $b$ or $c$. For instance, an algebraic model
\begin{equation}
 \sum_i m_i+\sum_{i<j}a_{ij}\,\llcol{i}{j}+\sum_{i<j}b_{ij}\,\llss{i}{j}~, 
\end{equation}
was written down recently~\cite{An:2020jix}. The parameters $m_i$, $a_{ij}$ and $b_{ij}$ are empirically adjusted from the ground states of ordinary mesons and baryons. Some earlier versions, restricted to the mass terms $m_i$ and the chromomagnetic terms $b_{ij}$  were used to speculated on light or heavy-light multiquarks. See, e.g., the refs.\ in \cite{An:2020jix}. 

Once the exercise of calculating the chromoelectric and chromomagnetic matrix elements is achieved, the masses of all heavy pentaquarks are estimated, and some of them lie below the threshold for dissociation into a meson and a baryon. Of course, this is just an indication. It remains to solve accurately the five-body problem for the configurations of interest. 
\section{Heavy hexaquarks}
\label{se:hexquarks}
Two years ago, a striking result was published by a lattice team~\cite{Junnarkar:2019equ}: several fully-heavy hexaquark states $QQQQQQ$ are found lighter than their lowest $QQQ+QQQ$ threshold. 
To check this conjecture, a careful six-body calculation was carried out, using a potential model of the type  \eqref{eq:pot1}-\eqref{eq:pot2}: no bound state was found~\cite{Richard:2020zxb}.  This analysis survives the changes of the details of the potential, provided the hexaquark and the baryons entering the threshold are estimated consistently. The debate between potential models and lattice simulations becomes even more serious, as a recent lattice estimate of the $\Omega_{ccc}\Omega_{ccc}$ system finds it in the unitary regime, i.e., at the edge of binding \cite{Lyu:2021qsh}. Further studies are undertaken~\cite{Huang:2020bmb}. 
\section{Improvement of potential models}
The above potential models are somewhat crude. There is, however, a reasonable hope that the parameters somehow include several effects in an effective way, so that the conclusions about the bound or unbound character of multiquarks are rather robust. Let us discuss two  of these corrections, which are part of the ambitious program outlined by Nathan Isgur and his collaborators: ``taking the `naive' and 'non-relativistic' out of the quark potential model''~\cite{Capstick:1986kw}. 
\subsection{Relativistic kinematics}
From the operator inequality $(\vec p^2+m^2)^{1/2}-m<\vec{p}^2/2m$, one infers that using a relativistic form of kinetic energy lowers the energy of quark systems, for a given interaction. 
Of course, this is just one out of several effects of relativity, but this is the most commonly investigated. If one compares this peculiar relativistic correction in mesons and tetraquarks, it is found that the decrease in energy is more pronounced in the former, so that the relativistic kinematics works against the binding of tetraquarks~\cite{Richard:2021lce}. 
\subsection{String potential}
\begin{figure}[ht!]
 \centering
\raisebox{.8cm}{ \includegraphics[width=.2\textwidth]{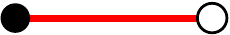}}\quad
 \includegraphics[width=.2\textwidth]{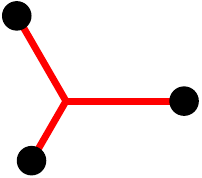}\quad
\raisebox{.2cm}{  \includegraphics[width=.2\textwidth]{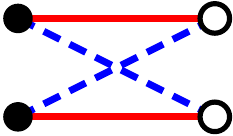}}\quad
 \raisebox{.2cm}{ \includegraphics[width=.2\textwidth]{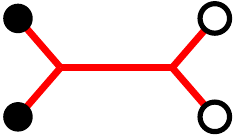}}
 \caption{String potential: mesons, baryons and tetraquarks (flip-flop and connected). This is the planar version. For a spatial view of the connected diagram, see, e.g., \cite{Ay:2009zp}.}
 \label{fig:string}
\end{figure}

One of the debatable properties of the potential \eqref{eq:pot1} is its pairwise character, especially for its linear part. Many years ago, it was suggested that the potential $a\,r$ between a quark and an antiquark corresponds to the energy of a gluon flux of constant section linking the two constituents, and that its generalization to baryon corresponds to  three fluxes linking each  quark to a common junction, whose location is adjusted to minimize the energy. See Fig.~\ref{fig:string}. This is the famous $Y$-shape interaction
\begin{equation}\label{eq:Y-shape}
 V=a\,\min_J (r_{1J}+r_{2J}+r_{3J})~,
\end{equation}
which evokes the Fermat-Torricelli problem. Some 3-body calculations have been done with this $Y$-shape potential 
\cite{1983AnPhy.150..267R,Dmitrasinovic:2009dy}, and the overall conclusion is that replacing the pairwise $a\sum r_{ij}/2$ by \eqref{eq:Y-shape} does not modify dramatically the phenomenology of the quark model of baryons. However, when one tackles the sector of tetraquarks and higher configurations, the potential becomes more attractive. For tetraquarks, the potential is the minimum over the  disconnected and connected configurations shown in Fig.~\ref{fig:string}. For the former, often called the flip-flop,  the minimum is taken of the two possible pairings.   
If one solves the four-body problem with the potential of Fig.~\ref{fig:string}, one finds a bound state, even for equal masses, provided that there is no antisymmetrization constraint \cite{Richard:2017vry}. Similar results were found for five- and six-body configurations \cite{Richard:2016eis,Richard:2020uan}. However, the flip-flop term, that largely dominates the connected one, requires that the color state can be freely rotated when the quarks move. Thus the string potential, if applied to, e.g., $cc\bar c\bar c$ requires further projections that spoiled the extra attraction. 
\section{Resonances}
It is a dream to discover a stable multiquark that decays only by Zweig-forbidden internal annihilation or by weak interaction. Nevertheless, the part of spectrum above the threshold is also extremely instructive.  The results of the past decades on $XYZ$ states and hidden-charm pentaquarks \cite{Brambilla:2019esw} correspond to resonances. They are often described as resulting from some hadron-hadron resonances. It has been shown that they can also be approached in the quark model, provided the few-body dynamics in the continuum is treated properly, for instance by real or complex scaling, as, e.g., in~\cite{Meng:2020knc} and refs.\ there. This effort will certainly be resumed, leading to a unified picture of bound and unbound multiquarks. 
\section{Outlook}
Within standard quark models, no bound multiquark state is found that contains solely heavy quarks $c$ or $b$. Therefore the discrepancy with respect to some lattice calculations remains to be understood. A warning is also set against the risk of producing artificially bound states by using unjustified approximations. For the resonances, some  methods borrowed form atomic and nuclear physics look rather promising. 

\subsection*{Acknowledgments}
I thank the organizers for maintaining this conference in spite of the present difficulties, Alfredo Valcarce and Javier Vijande for a long and friendly collaboration on the topics discussed in this contribution, and M.~Asghar for useful comments on the manuscript.  
%

%
\end{document}